\documentstyle[psfig]{mn2e}
\title[Radio QSOs at redshift 4]{A sample of radio-loud QSOs at redshift 
  $\sim$ 4}
\author[J. Holt et al]{J. Holt,$^{1,4}$
\thanks{Email: j.holt@sheffield.ac.uk}
C.R. Benn,$^1$
M. Vigotti,$^2$ 
M. Pedani,$^3$
R. Carballo,$^{5}$
J.I. Gonz\'{a}lez-Serrano,$^{6}$
\newauthor
K.-H. Mack,$^{7,2,8}$
B. Garc\'\i a$^{1}$ 
\\
$^1$Isaac Newton Group, Apartado 321, E-38700 Santa Cruz de La Palma, Spain\\
$^2$Istituto di Radioastronomia, CNR, via Gobetti 101, 
I-40129 Bologna, Italy\\
$^3$Centro Galileo Galilei, E-38700 Santa Cruz de La Palma, Spain \\
$^4$Department of Physics \& Astronomy, University of Sheffield, Hicks Building,
Sheffield S3 7RH, UK \\
$^5$Departamento de Matem\'atica Aplicada y C. Computaci\'on, 
Universidad de Cantabria,
E-39005 Santander, Spain\\
$^6$Instituto de F\'\i sica de Cantabria (CSIC-UC), Facultad de Ciencias,
E-39005 Santander, Spain\\
$^7$ASTRON/NFRA, Postbus 2, NL-7990 AA Dwingeloo, The Netherlands \\
$^8$Radioastronomisches Institut der Universit\"{a}t Bonn,
Auf dem H\"{u}gel 71, D-53121 Bonn, Germany \\
}
\begin{document}
\maketitle

\begin{abstract}
We obtained spectra of 60 red, starlike objects ($E<$ 18.8)
identified with FIRST radio sources, $S_{1.4GHz} >$ 1 mJy.
Eight are QSOs with redshift $z>$ 3.6.
% (4 of these with $z>$ 4.0). 
Combined with our pilot search (Benn et al 2002), our sample
of 121 candidates yields a total of
18 $z >$ 3.6 QSOs (10 of these with $z >$ 4.0).
8\% of 
candidates with $S_{1.4GHz}<$ 10 mJy, and 37\% of
candidates with $S_{1.4GHz}>$ 10 mJy are QSOs with $z >$ 3.6.
%FIRST sources, which are starlike on APM/POSS-I and on POSS-II, 
%with $O-E >$ 3, is 76\% complete over 7030 deg$^2$, i.e.
The surface density of $E <$ 18.8, $S_{1.4GHz} >$ 1mJy, 
$z>$ 4 QSOs is
0.003 deg$^{-2}$.
This is currently the {\it only}
well-defined sample of radio-loud QSOs at $z \approx$ 4
selected independently of radio spectral index.
The QSOs are highly luminous in the optical (8 have $M_B <$ -28, 
$q_0$ = 0.5, $H_0$ = 50 kms$^{-1}$Mpc$^{-1}$).
The SEDs are as varied as those seen in optical searches for high-redshift
QSOs, but the fraction of objects with weak (strongly self-absorbed)
Ly$\alpha$ emission is marginally 
higher (3 out of 18) than for high-redshift QSOs from SDSS (5 out of 96).
\end{abstract}


%========================================================================
\section{Introduction}
The evolution with redshift of the space density of QSOs places
strong constraints on the abundance of massive objects at the earliest
cosmological epochs.
High-redshift QSOs can be found efficiently by colour selection 
from very large samples (e.g. the Sloan Digital Sky Survey, Anderson et al
2001, see Benn et al 2002, hereafter Paper I, for a summary of previous
searches),
but space-density measurements based on such selection could be
biased by e.g. redshift-dependent dust extinction.

Any such bias is reduced when selecting in the radio, but searches for
high-redshift radio QSOs have so far concentrated on radio-bright
objects with flat radio spectra, which yields small samples with
relatively low surface density on the sky.
E.g. Snellen et al (2001) and Hook et al (2002)
sought red starlike optical counterparts of $S_{5GHz} >$ 30 mJy 
flat-spectrum radio sources,
and identified a total of 8 $z >$ 4 QSOs.
Here, we aim to identify high-redshift radio quasars without
any bias in radio spectral index, and with higher surface density
on the sky, 
through spectroscopy of red starlike counterparts of  
FIRST radio sources, with $S_{1.4GHz} >$ 1 mJy.
Combined with our pilot search
(Paper I), 
this yields 18 $z >$ 3.6 
radio-loud\footnote{All 18 QSOs are 
radio-loud according to the criterion of Gregg et al 1996, i.e. P$_{1.4 GHz}$
$>$ 25.5; see Table 1). } QSOs (9 previously known, mainly from
multi-colour optical searches).
This includes the largest sample to date of radio-selected
QSOs at z $>$ 4. 
%From this sample, Vigotti et al (2003) have
%measured the decline in space density of radio loud QSOs between z = 2
%and z = 4. [remove this sentence?? it's repeated in conclusions]

%==========================================================================
\begin{table*}
 \vbox to220mm{\vfill Landscape table to go here.
 \caption{}
 \vfill}
 \label{landtab}
\end{table*}

\begin{table*}
\begin{minipage}{170mm}
\caption{Candidates which are not $z >$ 3.6 QSOs}
\begin{tabular}{r r rrr rrl ll}
\hline
\multicolumn{1}{c}{RA}   & 
\multicolumn{1}{c}{Dec}   & 
\multicolumn{1}{c}{S$_{1.4}$}  & 
\multicolumn{1}{c}{R-O}    & 
\multicolumn{1}{c}{$E$} &
\multicolumn{1}{c}{$O-E$}&
\multicolumn{1}{c}{ID}&
\multicolumn{1}{c}{z}&
\multicolumn{1}{c}{$\sigma_z$}  &
\multicolumn{1}{c}{Notes}
\\

\multicolumn{1}{c}{J2000}   & 
\multicolumn{1}{c}{J2000}   & 
mJy     & 
\multicolumn{1}{c}{$\prime\prime$}
&     &     &   &  & &     \\
\multicolumn{1}{c}{(1)}  & 
\multicolumn{1}{c}{(2)}   & 
\multicolumn{1}{c}{(3)}&
\multicolumn{1}{c}{(4)}&
\multicolumn{1}{c}{(5)}&
\multicolumn{1}{c}{(6)}&
\multicolumn{1}{c}{(7)}&
\multicolumn{1}{c}{(8)}&
\multicolumn{1}{c}{(9)}&
\multicolumn{1}{c}{(10)} \\
\hline
\multicolumn{10}{l}{\bf Low-redshift QSOs and galaxies
(`?' = probable galaxy)} \\
\rule{0mm}{3.5mm}
 08 53 36.66& 17 43 47.9&  40.4&0.4&18.4&    3.2&Q&     &     &Jaunsen et al (1995), $z$ = 3.21 \\
 09 29  00.44& 16 28 06.3&   4.8&0.9&17.0&$>$ 4.7&?&     &     &                                  \\
 09 41 39.71& 35 32 33.5&   4.6&0.6&18.8&$>$ 2.9&?&     &     &                                  \\
 09 44 28.41&  06 41 44.7&   3.2&0.8&18.6&$>$ 3.1&?&     &     &                                  \\
10  01 42.41& 20 48 18.2&   2.2&0.5&18.6&$>$ 3.1&Q&     &     &Stocke et al (1991), BL Lac, $z$ = 0.346 \\
10  08 34.99& 35 51 23.4&   6.5&1.1&18.6&$>$ 3.1&?&     &     &                                  \\
10 13 08.38&  08 27 15.7&  28.5&0.4&18.7&$>$ 3.0&?&     &     &                                  \\
10 24 42.51& 56 11 25.1&   0.8&0.3&18.7&$>$ 3.0&?&     &     &                                  \\
10 25  06.45& 19 15 44.6&   2.0&0.7&18.6&$>$ 3.1&?&     &     &                                  \\
11 34 58.18& 12 32 21.2&   6.7&0.5&18.0&    3.3&?&     &     &                                  \\
12  02  09.51& 31 30 40.1&   5.7&1.0&18.7&$>$ 3.0&?&     &     &                                  \\
12  05  01.33& 14 28 30.7&  15.4&1.4&18.7&    3.2&?&     &     &                                  \\
12 20 16.87& 11 26 28.2&   2.3&0.4&18.1&    3.6&Q&1.893&0.010&CIII 1909, CIII 2326, MgII 2798 \AA  \\
12 39 32.76& -00 38 38.1&   1.3&0.6&18.4&$>$ 3.3&Q&     &     &Boyle et al (1990), BAL, $z$ = 2.18\\
12 39 56.76& 42 40 59.8&   3.2&0.4&18.8&$>$ 2.9&?&     &     &                                  \\
12 53 56.32&  00 52 42.9&  12.7&1.4&18.7&$>$ 3.0&?&     &     &                                  \\
12 55 04.57& 43 10 38.1&   3.2&0.2&18.8&$>$ 2.9&?&     &     &                                  \\
12 55 27.65& 55 18 19.0&   1.3&0.8&18.6&$>$ 3.1&?&     &     &                                  \\
13  00 38.85& -03 35 16.8&   1.4&0.6&17.4&    3.1&G&     &     &      
Extended on JKT image, no spectrum obtained
      \\
13  00 53.46& -03 11 28.2&   7.5&1.0&18.0&    3.0&G&     &     &                  \\
13  06 21.99&  03 36  06.3&   2.0&0.8&18.4&$>$ 3.3&?&     &     &                                  \\
13 10 43.44& 43 39  04.8&   4.7&0.7&18.6&$>$ 3.1&?&     &     &                                  \\
13 18 51.57& -00 53 23.4&   4.1&1.5&18.5&    3.6&?&     &     &                                  \\
13 32 29.08& 26 34 33.2&   1.3&0.3&18.8&$>$ 2.9&?&     &     &                                  \\
13 39 43.42& 40 34 25.4&   1.0&1.0&18.7&$>$ 3.0&?&     &     &                                  \\
13 43 14.20& 22  08 45.7&   1.5&0.4&18.4&$>$ 3.3&G&     &     &                  
Extended on JKT image, no spectrum obtained
\\
13 49 07.12& -02 23 24.9&   3.1&0.4&18.0&$>$ 3.7&?&     &     &                                  \\
14 13 59.17&  02 55 27.2&   1.1&0.3&17.9&    3.1&G&     &     &                                  \\
14 20 54.98& 21 10 29.6&   1.9&0.5&18.1&$>$ 3.6&G&     &     &                                  \\
14 28  01.12& 25 45 40.2&   3.3&0.7&18.7&$>$ 3.0&?&     &     &                                  \\
14 32 22.27& 48 34 42.6&  16.7&0.7&18.8&$>$ 2.9&S&0.1906    &0.0005    &
OIII 4959/5007, H$_\alpha$                                  \\
14 52 43.77&  01 27 33.1&  11.9&0.5&18.5&$>$ 3.2&?&     &     &                                  \\
14 53 46.71& 35 53 11.6&   1.3&0.9&18.8&$>$ 3.0&G&     &     &                                  \\
15 11  07.20& 11 45 58.8&   1.0&0.4&18.8&$>$ 2.9&G&     &     &                  
Extended on JKT image, no spectrum obtained
\\
15 34 22.24& 18 51 25.2&   1.5&0.3&18.4&    3.2&?&     &     &                                  \\
15 45 20.00& 10 28 55.7&   5.6&1.3&18.8&$>$ 3.0&G&     &     &                                  \\
16 40 06.14& 17 38 49.4&  15.6&1.0&18.2&    3.0&?&     &     &                                  \\
17 00 01.29& 19 29 31.6&   2.7&0.8&18.7&$>$ 3.0&G&     &     &                  
Extended on JKT image, no spectrum obtained
\\
\multicolumn{10}{l}{\bf Stars} \\
09 01 46.15& 15 02 21.8&   3.7&1.3&18.6&$>$ 3.1&*&     &     &                                  \\
09 09 53.20& 03 01 35.8&   1.5&0.5&18.6&$>$ 3.1&*&     &     &                                  \\
09 12 20.95& 14 18 48.3&   1.8&1.5&18.4&$>$ 3.3&*&     &     &                                  \\
09 17 33.46& 08 56 30.7&   2.1&1.4&17.9&    3.7&*&     &     &                                  \\
09 27 20.11& 14 25 49.8&  11.4&0.5&18.7&$>$ 3.0&*&     &     &                                  \\
09 47 34.27&-01 51 31.7&   7.5&0.8&18.4&    3.2&*&     &     &                                  \\
09 51 53.79& 00 52 05.3&   7.2&0.8&18.6&    3.3&*&     &     &                                  \\
11 07 19.77& 15 16 20.1&  14.3&1.2&18.0&    3.0&*&     &     &                                  \\
11 09 25.01& 12 10 30.9&  11.1&1.0&18.3&    3.1&*&     &     &                                  \\
11 22 09.69& 02 13 36.8&   2.3&1.2&17.9&$>$ 3.8&*&     &     &                                  \\
12 13 06.97& 13 43 07.3&  62.7&0.5&18.4&    3.1&*&     &     &                                  \\
13 26 58.60& 17 35 14.5&  14.8&1.4&17.2&    3.2&*&     &     &                                  \\
14 33 23.85& 37 39 08.9&  15.8&1.2&18.7&$>$ 3.0&*&     &     &                                  \\
14 49 00.23& 07 37 36.2&   1.3&1.3&18.5&$>$ 3.2&*&     &     &                                  \\
15 14 50.17& 06 36 05.3&   2.4&1.2&17.2&    3.3&*&     &     &                                  \\
15 16 43.79& 22 20 46.3&   4.0&1.5&17.8&    3.7&*&     &     &                                  \\
15 39 59.30&-02 15 09.7&  14.0&1.3&18.3&$>$ 3.4&*&     &     &                                  \\
16 03 05.11& 14 43 42.9& 285.3&1.4&17.5&    3.1&*&     &     &                                  \\
16 23 57.72& 23 50 12.0&   2.5&1.0&18.8&$>$ 3.0&*&     &     &                                  \\
16 41 53.13& 12 55 12.7&   1.8&0.6&17.8&    3.4&*&     &     &                                  \\
16 48 20.27& 50 39 04.0&  62.6&1.2&18.8&$>$ 2.9&*&     &     &                                  \\
\hline
\end{tabular}
Columns 1-6 and 8-10 are as in Table 1.  
Column 7 indicates spectral type (see caption of Fig. 3).
For 3 of the QSOs,
redshifts are available from the literature
(indicated); spectra were not obtained of
these.
\end{minipage}
\end{table*}

\section{Sample}
The selection procedure is a refinement of that
described in Paper I.
We
sought red, starlike optical identifications of the 722354 sources
in the July 2000 edition of the FIRST catalogue
of radio sources (7988 deg$^2$, 7$^h$ $<$ RA $<$ 17$^h$,
-5$^o$ $<$ dec $<$ 57$^o$, White et al 1997,
$S_{1.4GHz}$(peak) $>$ 1 mJy), satisfying the
following criteria:
\begin{enumerate}
\item starlike optical counterpart in the 
APM (Irwin et al 1994) catalogue of the POSS-I survey
(which includes 7030 deg$^2$ of
the FIRST survey), lying within 1.5 arcsec of the FIRST radio source;
\item POSS-I/APM $E <$ 18.8;
\item POSS-I/APM $O-E \ge$ 3, or no
measured $O$ mag (plate limit $O \approx$ 21.7);
\item starlike on POSS-II images (this excludes 80\% of candidates 
satisfying the above criteria);
\item red also in the Minnesota APS (Pennington et al 1993) catalogue
of the POSS-I plates, i.e. not omitted from the APM catalogue because
of confusion on the blue plate, see Paper I (this excludes
16\% of the candidates satisfying the above criteria).
\end{enumerate}
This yielded 294 candidates, of which we visually classified 194 
as `starlike' on POSS-II
and 100 as `possibly starlike'.

CCD images were obtained in good seeing of about 
one third of these, using
the Loiano 1.5-m telescope of the Astronomical Observatory
of Bologna. On these images, 25 of the 70 `starlike' on POSS-II,
and 28 of the 33 `possibly starlike' candidates were extended.
We therefore excluded from our sample
the 100 `possibly starlike' objects, implying an incompleteness of
100 $\times$ 5/33 = 15.2 stellar candidates. 

The final list of 194 candidates is therefore expected to
include 
45/70 $\times$ 194 = 124.7 truly starlike images, out of a total of
124.7 + 15.2 = 139.9 amongst the original list of 294 candidates.
Therefore, it includes 89\% (= 124.7 / 139.9) 
of the red QSO identifications of sources
in this edition of the FIRST catalogue.
The selection criteria differ from the preliminary criteria used
in Paper I only in 
going slightly deeper, i.e. $E <$ 18.8 rather than $E <$ 18.6.

%===========================================================================
\section{Observations and reduction}
\subsection{Optical Spectroscopy}
Observations of 54 of the 194 candidates were reported in
Paper I. 
The 55th object reported there, 1349+38, is actually
detected on the Minnesota APS scan
of the POSS-I blue plate
(i.e. it is not red), and thus does not meet selection criterion
(v) above.
Spectra were obtained of 60 more candidates using the
IDS spectrograph on the
Isaac Newton Telescope 
during near-photometric nights (occasional light cirrus) in 2001 May.
The spectrograph was used with the R150V grating, yielding
spectra with dispersion
6.5 \AA/pixel, usually centred at 6500 \AA.
For details of the observing and data-reduction, refer to
Paper I.
%The observing and data-reduction procedures were as discussed in
%Paper I.
The results are summarised in Tables 1 ($z >$ 3.6 QSOs) and 
2 (other objects).
Table 2 includes an additional 3 QSOs 
which were not observed, but whose redshifts were obtained
from the literature.
Spectra of the QSOs with redshift $z >$ 3.6 are shown in Figure 1.

To test the completeness of the search for $z >$ 4 QSOs,
we observed an additional 
28 candidates with 2.0 $< O-E <$ 3.0.  None had $z >$ 3.6.
The spectra of two objects with $z$ = 3.2, 3.4 are given
in Figure 2 (details included in Table 1).

In total, we have spectroscopic information for 117 of the 194
candidates with
$O-E >$ 3 (54 observed Paper I, 60 observed here, 3 redshifts
$z <$ 3.6 
from the literature).  In addition, we were able to classify 
4 objects as galaxies on the basis of JKT imaging (see below).
The colour-magnitude distribution of these 121 candidates is shown
in Figure 3.

The distribution of our 18 $z >$ 3.6 QSOs in radio flux density and 
$E$ mag
is compared
with that from other radio searches for high-redshift QSOs in Fig. 4.
In the optical,
these QSOs are highly luminous, most with 
$M_{AB}$(1450\AA) $<$ -27.

Of the 121 observed candidates, 18 (15\%) are QSOs with $z >$ 3.6,
and 10 of these have $z >$ 4.

\subsection{Optical imaging}
Some of the 194 candidates were imaged through a Harris $R$ filter
with the 1.0-m Jacobus Kapteyn Telescope (JKT)
in service time, in order to identify extended objects.
Spectra were later obtained of most of these candidates
at the INT (above), but the `G' (galaxy) 
classifications of 4 objects in Table 2  
%1300-03, 1343+22, 1511+11 and 1700+19
are based on JKT imaging alone.

In addition, Harris
$R-$ and $I-$band photometry was obtained with the JKT (see Table 1) for 
9 of the $z >$ 3.6 QSOs,
primarily 
to allow an independent check of the derivation of 
$M_{AB}$(1450\AA) absolute magnitudes (see Vigotti et al 2003).
Conditions were photometric, except for light cirrus during 
observation of 1639+43.  Otherwise, the estimated errors on the 
R, I mags are $<$ 0.05 mag.
For smooth continuum spectra, one would expect
POSS-I E and JKT R mags to differ by
$<$ 0.1 mag (Humphreys et al 1991), at a given epoch.
The observed $R-E$ values range -0.8 to 1.4 (median 0.7), 
mainly because the $R$ filter covers a wavelength range a factor
of 3 larger than does the $E$ filter, and includes
more of the depressed continuum blueward of Ly$\alpha$.
The range of $R-E$ values may also reflect
changes in QSO luminosity
during the 50 years between the 
POSS-I and JKT observations.

\subsection{Radio observations}
The 13 $z >$ 3.6 QSOs with $S_{1.4GHz} >$ 3 mJy were observed 
at 4.85 and 10 GHz, with the Effelsberg radio telescope, 
during the night of 2002 Jan 13.
The sources are all unresolved by the telescope beam, FWHM
143 arcsec and 69 arcsec at these two frequencies.
The sources were observed by scanning the telescope in right ascension
and declination, with scan speeds 30 arcmin and 10 arcmin per minute
respectively, scan lengths 15 arcmin and 7 arcmin, with the
total number of scans adjusted to the total flux density of each source.
The flux densities (scale of Baars et al 1977) and spectral indices
(between 1.4 and 4.85 GHz) are given in Table 1.
The median spectral index is $\alpha$ = -0.3 ($S_\nu \propto \nu^\alpha$).

% Fig 1
%
\begin{figure*}
\centering
\psfig{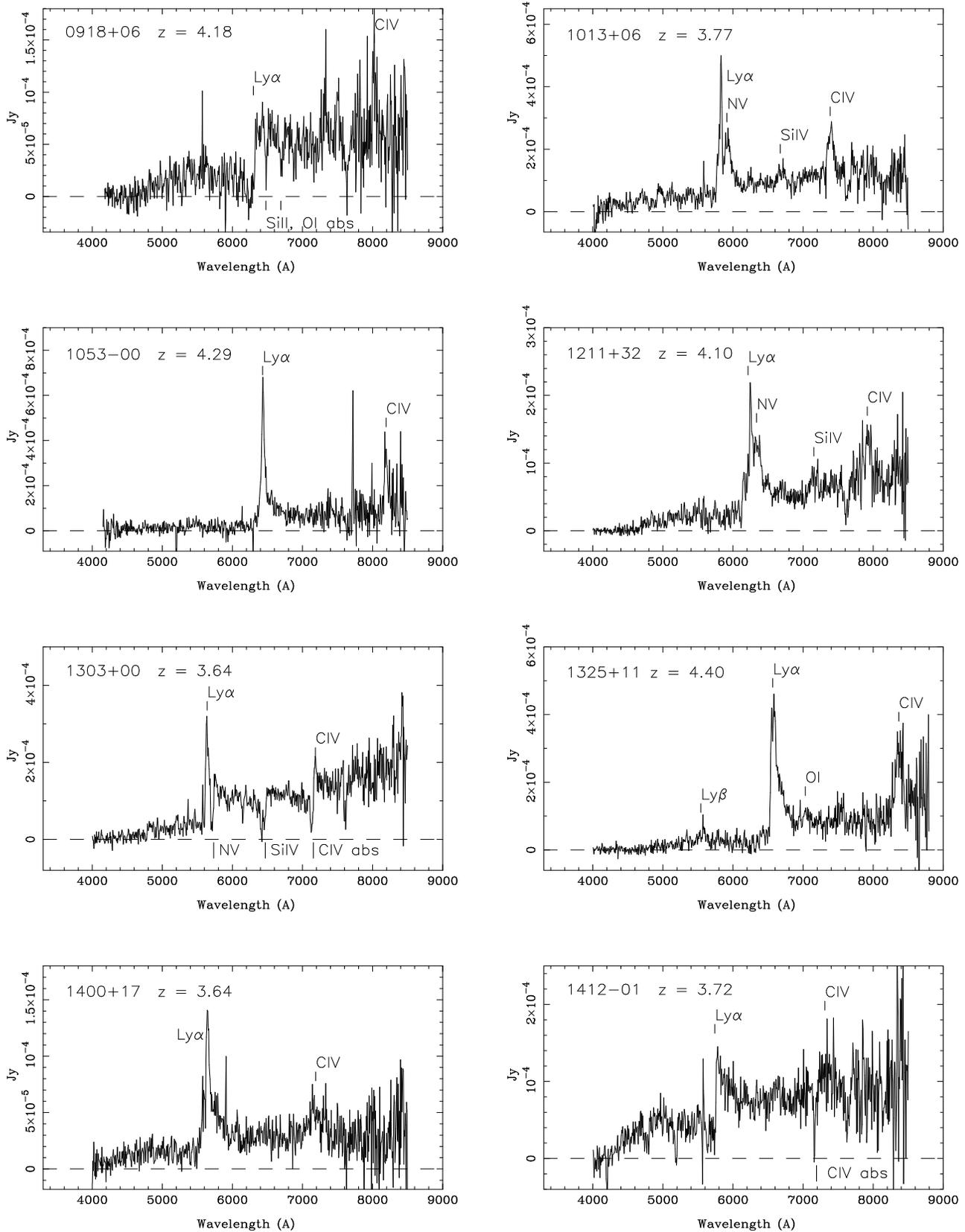}
\caption{
Spectra of the 8 $z >$ 3.6 QSOs amongst the 60 newly-observed
candidates (see Paper I for those previously discovered).
Spectral features are labelled at wavelengths corresponding to
the quoted redshift, assuming rest-frame wavelengths
in \AA\ of 1216 (Ly$\alpha$), 1240 (NV), 1302 
(OI/SiII blend), 1400 (SiIV/OIV] blend), 1549 (CIV).
The spectra have not been corrected for terrestrial atmospheric
absorption, notably at 7594 and 6867 \AA\ (A and B bands).
The ticks below the spectrum of 0918+06 indicate a
$z$ = 4.140 absorption system (1260 \AA\ SiII, 1302 \AA\ OI/SiII,
%1335 \AA\ CII, 
see also Snellen et al 2001), i.e. velocity
2330 km s$^{-1}$ relative to the QSO.
Those below the spectra of 1303+00 and 1412-01 indicate absorption 
at $z$ = 3.600 and 3.624 respectively,
i.e. outflow velocities 2600 and 6160 km s$^{-1}$.
}
\end{figure*}

%
% Fig 2
%
\begin{figure*}
\centering
\psfig{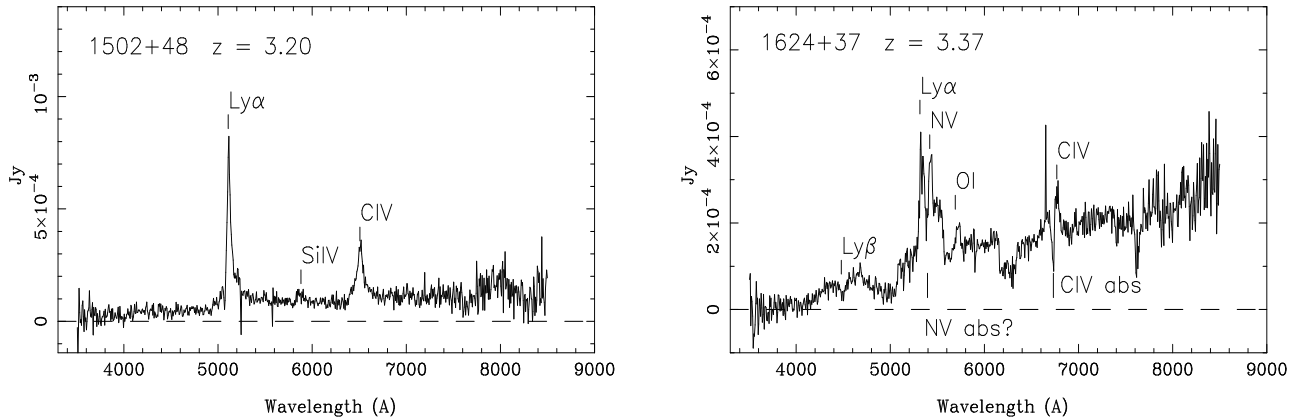}
\caption{
Spectra of 2 $z >$ 3 QSOs discovered when observing a sample of
bluer, 2.0 $< O-E <$ 3.0, candidates (i.e. not included in the main
sample of 121 observed candidates).
The ticks below the spectrum of 1624+37 indicate an absorption system
at $z$ = 3.327, i.e. velocity 2970 km s$^{-1}$ relative to the QSO.
This QSO also exhibits a CIV broad absorption line (BAL) with outflow
velocity up to 29000 km s$^{-1}$.
}
\end{figure*}

% Fig 3
%
\begin{figure*}
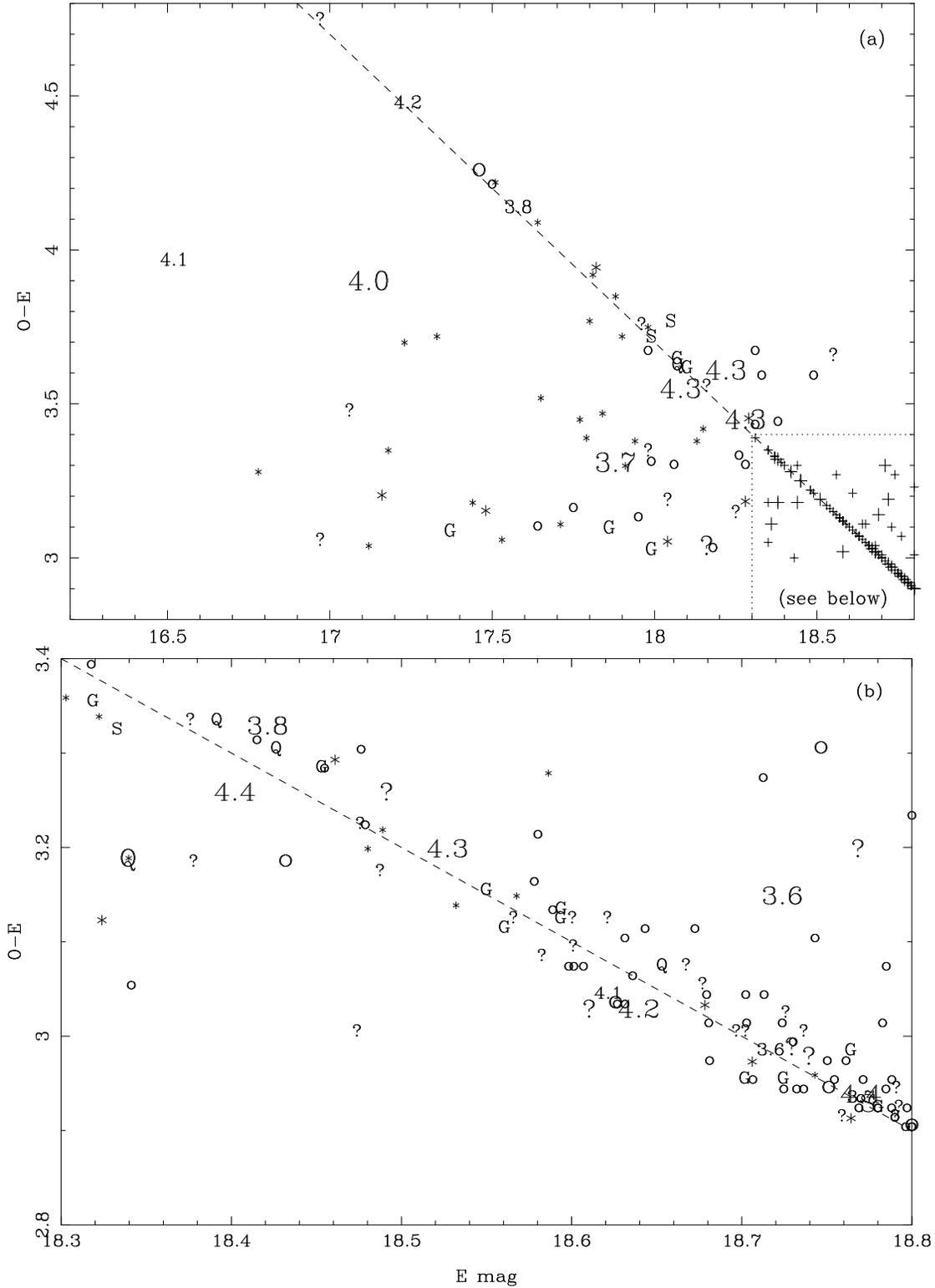

%\begin{minipage}{17cm}
%\centering
\centerline{\psfig{file=Holt-fig3a.ps,height=10.2cm,width=14.4cm,angle=270}}
\centerline{\psfig{file=Holt-fig3b.ps,height=10.2cm,width=14.7cm,angle=270}}
\caption{
(a) Distribution in colour and magnitude of the 194 starlike
candidates.
(b) Expanded view of the lower-right corner of Fig. 3a.
The symbols for spectral type (121 objects) are: number = redshift of QSO with
$z >$ 3.6, 
`Q' = QSO with $z <$ 3.6, 
`S' = emission-line galaxy (starburst or Seyfert),
`G' = radio galaxy, 
`?' = probable radio galaxy (definitely not high-redshift QSO, but might be 
a star), 
`*' = star.
Large font indicates $S_{1.4GHz} >$ 10 mJy;
small font $S_{1.4GHz} <$ 10 mJy.
Objects
with no $O$ magnitude have been plotted at the nominal $O$ plate limit
(dashed line) with $O-E$ = 21.7 - $E$.
The unobserved 73 = 194 - 121 objects are plotted as circles.
In Figure 3a, the positions of a few points have been adjusted to reduce
overlap of labels.  In Figure 3b, the E mags have been randomised $\pm$
0.05 mag for the same reason.
}
%\end{minipage}
\end{figure*}

%-----------------------------------------------------------------------------

\section{Results}
\begin{table*}
\label{Table 3}
\caption{Fractions of candidates which are QSOs with $z >$ 3.6}
\centerline{\begin{tabular}{c rrrrr} \hline
S$_{1.4GHz}$&
\multicolumn{1}{c}{E $<$ 17.8}&
\multicolumn{1}{c}{17.8 $<$ E $<$ 18.3}&
\multicolumn{1}{c}{E $>$ 18.3}&
\multicolumn{1}{c}{All} & 
\multicolumn{1}{c}{All \%}\\ 
(1)&
\multicolumn{1}{c}{(2)}&
\multicolumn{1}{c}{(3)}&
\multicolumn{1}{c}{(4)}&
\multicolumn{1}{c}{(5)}&
\multicolumn{1}{c}{(6)}
\\
\hline
$<$ 3  & 4/16& 0/13& 1/28 & 5/57 & 9\\
3 - 10 &  0/7&  0/8& 2/19 & 2/34 & 6\\
$>$ 10 &  1/3&  4/9& 6/18 & 11/30& 37\\
All    & 5/26& 4/30& 9/65 & 18/121 & 15 \\
All \% & 19 & 13 & 14 & 15 \\
\hline
\end{tabular}}
Columns are: (1) FIRST 1.4-GHz integrated flux density (mJy), 
(2-4) number of $z >$ 3.6 QSOs / total number of candidates, for each
range of POSS-I $E$
magnitude, (5) sum over all ranges of $E$, and (6) sum
expressed as a percentage.
\end{table*}

\subsection{High-redshift QSOs}
The sample comprises 18 $z >$ 3.6 QSOs (Table 1).
Below we derive the surface density 
on the sky, and we comment on the spectra of individual
objects.

\subsubsection{Surface density}
The fraction of  
high-redshift QSOs amongst the candidates does not depend on optical
magnitude, but depends strongly on radio flux density
(Table 3):
8\% for $S_{1.4GHz} <$ 10 mJy, 37\% for 
$S_{1.4GHz} >$ 10mJy.
73 of the 194 candidates have not been
observed, 
67 with $S_{1.4GHz} <$ 10 mJy,
6 with $S_{1.4GHz} >$ 10 mJy.
On the basis of the above high-redshift-QSO fractions, 
we expect to have missed 5.95 $z >$ 3.6 QSOs in the sample of
194 candidates, i.e. the completeness of our search for candidates
satisfying our selection criteria is
(18 / (18+5.95)) = 75\% over the 7030 deg$^2$ overlap between 
FIRST and the APM/POSS-I catalogues, within 7$^h$ $<$ RA $<$ 17$^h$.
Multiplying by the 89\% completeness due to filtering out
`possibly-starlike' candidates (Section 2), our search is 67\% complete,
% 66.9\% to be exact
i.e. the surface density of $z >$ 4 QSOs 
(10 QSOs $E <$ 18.8, $S_{1.4GHz} >$ 1 mJy)
found with this technique is 0.0021 deg$^{-2}$.

This is about twice that of recent 
searches for high-redshift QSOs amongst the
red starlike optical counterparts of
(radio-brighter) flat-spectrum sources in the northern
(Snellen et al 2001; 4 $z>$ 4, $E <$ 19, 0.0006 deg$^{-2}$) and 
southern (Hook et al 2002; 4 $z>$ 4, $S_{5GHz} >$ 72 mJy,
$R <$ 22.5, 0.001 deg$^{-2}$) hemispheres.
The surface densities measured for the FIRST sources are higher
in part because of the much fainter radio flux-density limit,
and in part because steep-spectrum sources are not excluded.

In the sample of Hook et al (2002), the ratio of the numbers of
$E <$ 21 and $E <$ 19 $z>$ 4 QSOs is 7 / 4 = 1.8.
Combining this with our above result, we predict the surface 
density of $E<$ 21, $S_{1.4GHz} >$ 1 mJy QSOs with $z >$ 4  
to be 0.004 QSOs deg$^{-2}$.

The surface-density calculations above account for
incompleteness  due   to rejection  of  candidates
not considered definitely starlike on POSS-II, and due to only
121 out of 194
candidates being observed.
In calculating the overall completeness of a search for $z>$ 4 QSOs
with $S_{1.4GHz} >$ 1 mJy,
% $E <$ 18.8, $S_{1.4GHz} >$ 1 mJy, 
one must also take into account:
\begin{enumerate}
\item the completeness of the FIRST catalogue at low flux densities,
  83\% for the flux-density distribution of Table 1 (Prandoni et al 2001);
\item the completeness of the APM catalogue of POSS-I, 84\% (see Vigotti et al 2003, 
      section 2.3);
\item that the radio-optical separations of some QSOs may exceed
1.5 arcsec due to measurement errors, completeness 99\%, given combined
radio-optical rms 0.5 arscec;
\item  that some QSOs may be missed because they do not coincide with
peaks of radio emission (e.g.
near the mid-points of double radio sources), completeness 98\% 
(see Vigotti et al 2003, section 2.3);
\item that some $z>$ 4 QSOs may have $O-E <$ 3, completeness $\sim$ 100\%
(Section 3.1, see also Vigotti et al 2003, figure 1).
\end{enumerate}
The completeness due to these five factors combined 
is 0.68, i.e. the corrected
surface densities of $z >$ 4 QSOs with $S_{1.4GHz} >$ 1 mJy
is 0.0031 deg$^{-2}$ for $E <$ 18.8, and
is 0.006 deg$^{-2}$ for $E <$ 21.

Since $\sim$ 10\% of QSOs are radio-loud 
(e.g. Sramek \& Weedman 1980; and Petric et al 2003 for
consistency at $z \sim$ 5), we expect the density
of radio-quiet + radio loud $z >$ 4 QSOs with $E <$ 21 to be
$\sim$ 0.06 deg$^{-2}$.
E.g. Kennefick et al (1995a,b) find 0.03 deg$^{-2}$ with $r <$ 20.
The SDSS search (e.g. Anderson et al 2001) reaches a higher surface density,
$\sim$ 0.1 deg$^{-2}$ because the $i$ and $z$ bands are also used, i.e. no
effective limit in $r$ band.

\subsubsection{Ly$\alpha$ emission}
The rest-frame Ly$\alpha$+NV  equivalent widths of the 18
$z >$ 3.6 QSOs are given in Table 1.
The values are approximate, rms $\sim$ 30\%, 
due to the difficulty of estimating 
the slope of the underlying continuum
(see e.g. Schneider 1991), but the median value of 75 \AA\
is similar to 
that reported elsewhere for
$z \sim$ 4 QSOs, $\sim$ 70 \AA\ (Fan et al 2001, Constantin
et al 2002).
However, 3 of our 18 radio-selected QSOs have 
rest-frame equivalent widths $<$ 25 \AA\ :
0831+52 ($<$ 10 \AA), 0918+06 ($<$ 10 \AA)
and 1639+40 ($\sim$ 15 \AA), presumably because the lines are
heavily self-absorbed.
By comparison, only 5 out of 93 optically-selected SDSS QSOs
$z >$ 3.6
(from Fan et al 1999a, Fan et al 2000, Fan et al 2001,
Schneider et al 2001, Anderson et al 2001) exhibit such small
Ly$\alpha$ equivalent widths
(but see Fan et al 1999b for an example of a $z$ = 4.62 QSO with 
no emission lines at all).
The difference between the two samples is weakly significant.
Our sample of QSOs and that from SDSS have similar median redshift,
so the difference cannot be ascribed to a difference in epoch
(e.g. at $z >$ 5.7, Fan et al 2003 found that 1 out of 3 
QSOs has Ly$\alpha$ equivalent width $\leq$ 25 \AA). 
A difference might also arise due to the inverse correlation
between emission-line equivalent width and 
optical luminosity known at low redshift, the Baldwin effect
(Baldwin et al 1989).  
The optical luminosities of our
QSOs are indeed $\sim$ 2 mags higher than those from SDSS,
but Constantin et al (2002) found the Baldwin  effect
to be weak or absent for QSOs with $z >$ 4.
Finally, the difference might be connected with the fact that our QSOs host a 
radio source, although
previous studies (e.g. Corbin 1992) have found that emission-line
strengths do not depend on whether a QSO is radio loud or quiet.
If one combines
our radio-selected sample (18 QSOs) with the radio-selected
(but radio brighter)
samples of Hook et al (2002; 13 QSOs $z >$ 3.6 with spectra) 
and of Snellen et al (2001;
4 QSOs $z >$ 3.6, one already included in our sample), 
the fraction with 
Ly$\alpha$ equivalent widths $<$ 25 \AA\ reduces to
3 out of 34, i.e. not significantly different from SDSS.

High-redshift QSOs are also characterised by a drop in the continuum
across the Ly$\alpha$ line, from red to blue, due to absorption by
neutral hydrogen along the line of sight.
The measured values of 
the opacity $D_A$ = (1 - $f_1$ / $f_2$), where $f_1$ and $f_2$ are the 
intensities in the continuum regions
1050 - 1170 \AA\ and 1250 - 1350 \AA\ respectively
(Oke \& Korycansky 1982), are given in column 16 of Table 1.
The error in measured
$D_A$ is $\sim$ 10\%.
The median $D_A$ for the 8 QSOs with 3.6 $< z <$ 4.0 is 0.50,
while that for the 10 QSOs with 4.0 $< z <$ 4.4 is 0.62,
consistent with the trend with redshift observed by Kennefick et al
(1995b, their fig. 5) for optically-selected QSOs.

\subsubsection{Narrow absorption lines}
The QSOs 0831+52, 0918+06,
0941+51, 1303+00 and 1412-01 show metal-line
absorption features at velocities relative to the QSO
ranging 1200 to 7000 km s$^{-1}$.
%, and with FWHM ranging 1500 to 5000 km s
QSOs 0831+41, 0918+06, 0941+51 and 1303+00 exhibit absorption 
at velocity separations below 3000 km s $^{-1}$ (i.e. probably
associated with the QSO).
0831+41 and 1412-01 exhibit
absorption systems  at higher velocities.
Metal absorption lines are not visible in our spectrum of 0747+27,
but in a high-resolution spectrum obtained by Richards et al (2002),
at least 14 independent CIV absorption systems are detected.

%
% Fig 4
%
\begin{figure}
\centering
\psfig{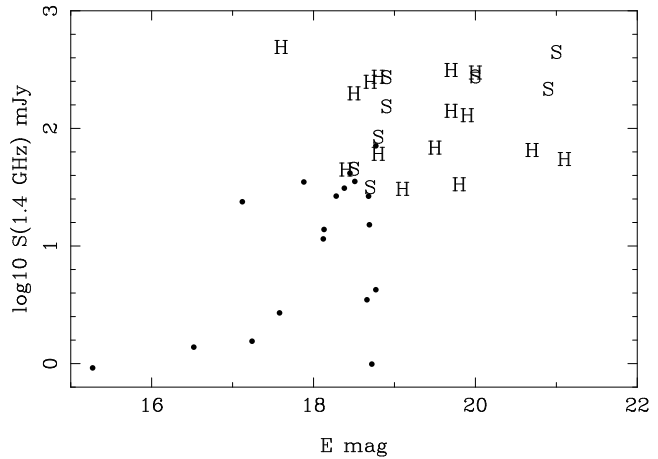}
\caption{
Distribution in $S_{1.4GHz}$ and $E$ mag of 
$z >$ 3.6 radio-selected QSOs.
$\bullet$ = FIRST QSOs (Table 1), S = Snellen et al (2001),
H = Hook et al (2002).
See also Isaak et al (2002) for a list of 17 $z >$ 4 optically- and
X-ray-selected QSOs with known radio counterparts.
}
\end{figure}

\subsubsection{BAL QSOs}
One of the 18 $z >$ 3.6 QSOs,
0831+41, is classified as a
 broad-absorption-line (BAL) 
QSO by Irwin et al (1998). 
Fan et al (2001) found 2 classical BALs in their sample of 39 high-z
QSOs.  The BAL fraction in both our sample and that of
Fan et al is consistent with 
the fraction 10\% found for
low-redshift optically-selected QSOs 
(Weymann et al 1991).

1624+37 ($z$ = 3.4, i.e. not included in the sample of 18 $z >$ 3.6 QSOs,
see Table 1) 
is a new BAL QSO. 
The CIV absorption feature has FWHM $\sim$
9000 km s$^{-1}$ and is detached by 20000 km s$^{-1}$ from the peak of the
emission.  
It has an unusually sharp high-velocity cutoff.

\subsection{Low-redshift objects}
The remaining candidates (Table 2) are a
mixture of low-redshift QSOs, emission-line galaxies, absorption-line
galaxies, and stars. The QSOs and galaxies are mostly true identifications,
the stars are
mostly misidentifications, as is implied by the distributions 
of radio-optical separations in Table 2 (although the higher values 
for stars might also be attributed to proper motion between the epochs
of POSS-I and the the FIRST survey).
The distribution of radio flux-densities of the 
sources apparently identified with stars is statistically indistinguishable
from that of the FIRST catalogue as a whole.
We obtained deep images at the JKT, $R <$ 24, of two 
M stars lying close to $S_{1.4GHz} \sim$ 100 mJy sources 
(0741+33, $E$ = 18.4, and 1353+34, $E$ = 17.8 from Paper I), 
but a host galaxy is not detected in either case,
so we cannot reject
the hypothesis that these are true radio stars.

A `?' classification in Table 2 indicates that there was insufficient
signal-to-noise to distinguish between galaxy and M-star spectra, but
that the spectrum is not consistent with that of a $z \sim$ 4 QSO
(i.e. we expect there to be $<\sim$ 1 such QSO hidden amongst the `?'
objects).

%===========================================================================
\section{Conclusions}
We identified 194 high-redshift QSO candidates which:
\begin{enumerate}
\item coincide with FIRST radio sources $S_{1.4GHz} >$ 1 mJy;
\item are classified starlike on APM (POSS-I) and on POSS-II;
\item have APM $E <$ 18.8 and $O-E \geq$ 3.0, or are invisible on
the $O$ plate.
\end{enumerate}
The sample covers an area of 7030 deg$^2$.
We have observed 
121 of the candidates and find  
18 to be QSOs with $z >$ 3.6, 10 of these with $z >$ 4
(Benn et al 2002, and this paper).
We estimate that we have found 75\% of the high-redshift QSOs
present amongst the 194 candidates.
The surface density of $z >$ 4 QSOs with $E <$ 18.8, $S_{1.4GHz} >$ 1 mJy, 
is 0.0031 deg$^{-2}$.

This is currently the {\it only}
well-defined sample of $z \approx $ 4 radio QSOs selected
independently of radio spectral index.
Vigotti et al (2003) use
a subsample of 13 of these QSOs,
with $z>$ 3.8, $M_{AB}$(1450 \AA) $>$ -26.9 and log
$P_{1.4 GHz}$(W Hz$^{-1}$) $>$ 25.7, to measure the  space density of
QSOs at $z$ = 4, and determine the change in space density
between $z$ = 2 and $z$ = 4.
%and to analyse its decline between $z$ = 2 and 
%$z$ = 4, using for $z$ = 2 a sample drawn
%from the  literature (White  et al.  1997, FBQS;  Boyle et
%al. 2000, 2dF).

These QSOs are highly luminous in the optical (8 have $M_B <$ -28,
$q_0$ = 0.5, $H_0$ = 50 kms$^{-1}$Mpc$^{-1}$).
The SEDs are remarkably varied.
They include:
3 QSOs with very weak Ly$\alpha$ (0831+52, 0918+06, 1639+40), 
one with an unusually-high density of CIV absorption systems
(0747+27, Richards et al 2002), 
one with a probable DLA (0941+51), 
the luminous and much-studied 0831+52 (Irwin et al 1998),
and a QSO with narrow Ly$\alpha$ of high equivalent width (1309+57).  
1624+37 (outside the sample) is an unusual BAL QSO.

\vspace{3mm}
{\bf Acknowledgments}\\
JH was a 1-year placement student at
ING during 2000-2001.
RC and JIGS acknowledge financial support from DGES project
PB98-0409-C02-02.
CRB, MV, RC and JIGS acknowledge financial support
from the Spanish Ministerio de Ciencia
y Tecnologia under project AYA2002-03326.
KHM was supported through a European Community Marie Curie fellowship.
The Isaac Newton and the Jacobus Kapteyn 
Telescopes are operated on the
island of La Palma by the Isaac Newton Group in the Spanish 
Observatorio del Roque de los Muchachos of the
Instituto de Astrofisica de Canarias.
%The NASA/IPAC Extragalactic Database (NED)
%is operated by the Jet Propulsion Laboratory, California
%Institute of Technology, under contract with NASA.
The 100-m Effelsberg radio telescope is operated by the 
Max-Planck-Institut f\"ur Radioastronomie. 
The APS Catalog of POSS I (http://aps.umn.edu) is supported by
NASA and the University of Minnesota.
We are grateful to the anonymous referee for helpful suggestions.

\end{document}